*Research Article*

# Filling Factors of Sunspots in SODISM Images

**Amro F. Alasta[1, \*], Abdrazag Algamudi[1], Fatma Almesrati[1], Mustapha Meftah[2] and Rami Qahwaji[1]**

[1]Electrical Engineering and Computer Science, University of Bradford, U.K
Amr_hard@yahoo.com; gamudi@yahoo.com; enas199990@yahoo.com; R.S.R.Qahwaji@bradford.ac.uk
[2] French National Centre for Scientific Research
Mustapha.Meftah@latmos.ipsl.fr
*Correspondence: Amr_hard@yahoo.com

**Abstract:** The calculated filling factors (FFs) for a feature reflect the fraction of the solar disc covered by that feature, and the assignment of reference synthetic spectra. In this paper, the FFs, specified as a function of radial position on the solar disc, are computed for each image in a tabular form. The filling factor (FF) is an important parameter and is defined as the fraction of area in a pixel covered with the magnetic field, whereas the rest of the area in the pixel is field-free. However, this does not provide extensive information about the experiments conducted on tens or hundreds of such images. This is the first time that filling factors for SODISM images have been catalogued in tabular formation. This paper presents a new method that provides the means to detect sunspots on full-disk solar images recorded by the Solar Diameter Imager and Surface Mapper (SODISM) on the PICARD satellite. The method is a totally automated detection process that achieves a sunspot recognition rate of 97.6%. The number of sunspots detected by this method strongly agrees with the NOAA catalogue. The sunspot areas calculated by this method have a 99% correlation with SOHO over the same period, and thus help to calculate the filling factor for wavelength (W.L.) 607nm.

## 1. Introduction

Observing the solar disk and detecting feature activities such as Sunspots is a core function of satellites such as PICARD. PICARD is a solar-terrestrial microsatellite mission launched in June 2010, with multi-institutional and international cooperation. Its overall objective is to monitor the solar diameter, the differential rotation, the total solar irradiance (simultaneous measurement of the absolute total and spectral solar irradiance), and to study the long-term nature of their inter-relations. The PICARD payload is composed of a an imaging telescope - the Solar Diameter Imager and Surface Mapper (SODISM) - which measures and records the solar diameter and shape to an accuracy of a few milliarcseconds and perfoms helio-seismologic observations to probe the solar interior.

SODISM (Solar Diameter Imager and Surface Mapper), is one of its on-board instruments. It has been developed for solar astrometry and helioseismology observations. SODISM allows us to measure the solar diameter and shape with an accuracy of few milliarcseconds and to perform helioseismologic observations to probe the solar interio[1]. The SODISM telescope observes the sun in five wavelengths, which are centred at 215.0, 393.37, 535.7, 607.1, and 782.2nm, he provided different image qualities depending on the wavelength[2].

Although SODSIM images deteriorate over time, automatic detection of sunspots using SODISM images is possible if appropriate pre-processing procedures are applied. The combination of solar irradiation and instrumental contamination significantly impacts on the quality of SODISM images, and causes degradation[3]. The W.L. 215nm channel lost more than 90% of its normalized intensity, and W.L. 393nm lost about 80% [4]; showing a pronounced degradation in the UV channels, as illustrated in Figure 1. The degradation arises due to the polymerisation of contaminants on the front window or on the other optical elements under the solar UV exposure. Meanwhile, the visible and near infrared channels present a temporal oscillation but remain relatively constant [3].

According to Figure 1, and with the exception of the 215nm W.L., all other wavelengths can be used to detect Sunspots. However, this paper investigates only 607 nm band images, which are available at level LB[1] (level B1 data products include a number of corrections for instrument issues). In total, approximately 250 images were downloaded, from 22th September 2010 to 4th January 2014. The format of these files was FITS, and each image has a size of 2048× 2048 pixels.

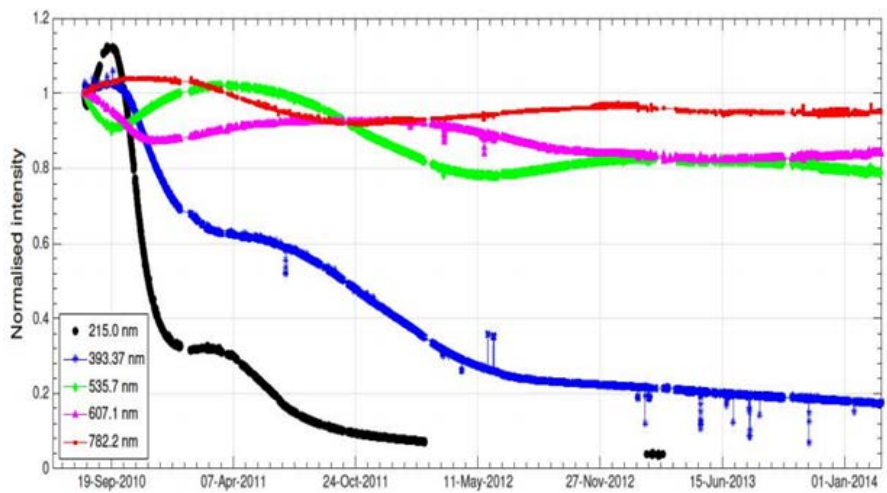

**Figure 1.** Normalized time series of integrated intensity of PICARD during its mission[4]

The detection and classification of sunspots is an important aspect in the monitoring and prediction of solar activity. The automated detection of sunspots and other solar features is valuable because, as this study confirms, it provides a robust, fast and accurate means of detection[5]. To detect solar features, researchers have used observatory images from the SOHO and SDO satellites, but the relative under use of images from SODISM has prompted an interest in working with SODISM images.

---

[1] http://picard.projet.latmos.ipsl.fr/files.php

This paper is an extension of our previous studies investigating the detection of sunspots using SODISM images. The paper presented in August 2018 at the IEEE International Conference on Computing, Electronics and Communications Engineering, has been especially helpful because research on the 'Identification of Sunspots on SODISM Full-Disk Solar Images' also compared the results with images from the NOAA catalogue[6]. This paper provides the following contributions:

- The provision of an automated method to detect Sunspots from SODISM images with verification for W.L. 607.
- The provision of a filling factor and comparisons with SOHO images in the same time period.
- A comparison with the NOAA catalogue.
- New method of summarising hundreds of experiments in a more concise manner presented to characterise SODISM data.

This paper is organized as follows: Section 2 surveys and summarizes the literature; Section 3 describes the pre-processing approaches applied; Section 4 shows detection and verification of Sunspots, section 5 illustrates the degree of accuracy between the NOAA and the proposed method; Section 6 provides the filling factor computations for the SODISM and SOHO findings, which presents some experimental results; Section 7 provides the statistical method to summarize the results of the SODSIM catalogue. Finally, the conclusions and results are presented in Section 8.

**2. Literature survey**

In order to discover its solar radius and centre, it is compulsory to detect the boundary limb of the solar disk before applying the segmentation features[7]. Once this has been established, the interior features can be analysed, such as Sunspots, which are dark and sometimes irregularly shaped local structures on the solar disk. There are three main approaches to segmentation [7]: Boundary-based, Region-based and Thresholding.

Of the three approaches, Thresholding is the simplest and quickest method[5]. However, the non-uniform brightness of the background solar disk makes the global thresholding of the solar disk an impractical solution. Nevertheless, this can be modified and corrected by normalizing the image brightness in a pre-processing step[6]. Furthermore, some background regions of the solar disk in some images have different contrasts and could be darker than some sunspots in other regions. However, global thresholding of the solar disk is not a practical solution because of the non-uniform brightness of the background solar disk caused by factors such as the limb darkening effect due to varying absorption of light in different thicknesses of the solar atmosphere traversed by rays from the sun to the detector[7].

Zharkov *et al.*[10] have summarized and evaluated fully-automated, manual and semi-automated feature recognition techniques and then applied them to different solar features. For example, in 2008, Curto *et al.*[8] provided a fully automated recognition approach to detect sunspots by using morphological operators. These recognition techniques[8] detect the boundaries between regions by looking for discontinuities in grey levels. Gauss smoothing and a Sobel gradient are applied to detect contours using operators that are sensitive to meaningful discontinuities in intensity levels. However, problems arise because these various applications produce unsatisfactory results; to counter this, post-processing operations are applied.

The previous methods applied to SODISM images to detect sunspots are summarized in the following studies:

Ashamari *et al.* introduced internal work2[11], which is a method for detecting sunspots from SODISM images (at a 535nm band). They first applied a Wavelet Harr filter to remove noise from the image and then used a band-pass filter to remove limb darkening. Finally, Gaussian smoothing was applied to remove isolated noisy pixels. Their results were excellent and the correlation coefficient between SOHO and SODISM images was found to be 0.98.

In comparison, in 2016 Meftah *et al.*[1] applied a similar method to Curto *et al.*[8] on SODISM 393nm data to detect sunspots and bright features. The steps of their method are as follows which produce results that reflect the same accuracy as manual thresholding:

- Firstly, apply pre-processing on L1 SODISM data in order to obtain SODISM images with a full contrast. To reduce noise, a Median filter is applied.
- Secondly, morphological processing is applied, consisting of a top-hat operation for Sunspot detection.
- Thirdly, the Otsu threshold for the segmentation of sunspots is applied.

For the detection of bright regions, morphological processing is performed, which consists of a bottom-hat (complementing the top-hat) operation. However, if the number of detected faculae is not coherent, an iterative procedure is launched; this starts from a reduced threshold and increases gradually until the number of detected faculae corresponds to a fixed interval. The main disadvantage of this method is the length of time it requires.

In 2017, Alasta *et al.*[7] applied SODISM data to a 535nm W.L. and their methods were as follows:

- Firstly, determine the solar disk and record its radius and centre information.
- Secondly, convert the image scale from a signed 32 bit, to an unsigned 8 bit.
- Thirdly, use Kuwahara and À Trous filters to remove noise and other unwanted features.
- Correct any pixels that are brightness outliers, and apply a Band pass filter to display the sunspots on a normalized background.
- Finally, apply a Threshold to obtain a mask image and determine the sunspot locations.

The results of this method were compared with the SOHO filling factor and the correlation coefficient between the two data sets was 98.5%.

In 2018, A. Alasta *et al.*[12] used wavelet domain to improve the resolution of SODISM images. This operation decomposes the original image into four lower resolution sub-bands, referred to as low-low (LL), low-high (LH), high-low (HL) and high-high (HH) as shown in Figure 2. The latter three sub-bands occupy the upper frequency spectrum of the original image and the resolution enhancement technique is applied here in this application[13]. Wavelet-based methods do this by enhancing image resolution through estimating the high frequency information from the given image. Such methods are based on the idea that the image to be enhanced is a low-frequency sub-band among wavelet-transformed sub-bands and the aim is to estimate the corresponding high-frequency sub-bands, so that an inverse wavelet transform can then be performed to obtain an image with enhanced resolution. The input image to be enhanced is regarded as a low-frequency sub-band in the context of an IDWT.

---

2 https://projects.pmodwrc.ch/solid/index.php/links/10-news-archive/31-deliverables

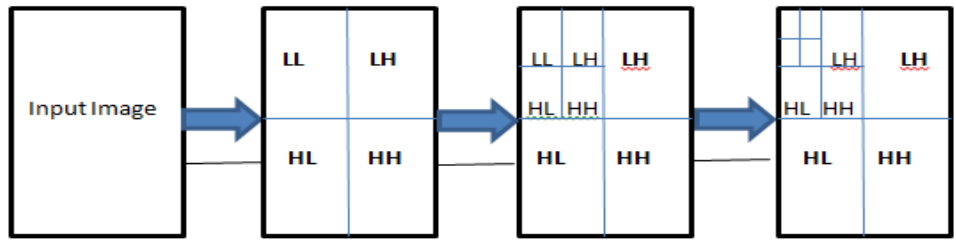

**Figure 2**. Applying the LL, LH, HL, and HH sub-band DWT three times on the lowest sub-band starting from the input image

In 2018, A. Alasta *et al.*[6] applied SODISM data to a W.L. 607nm to detect sunspots and compared the results with the NOAA catalogue; their methods detect the border of solar disks and then detect sunspot features. This method is completely automated, which makes it easy to apply to very large data images. The sunspot detection correlation coefficient is 0.99 with the SOHO satellite in the same time period.

## 3. Pre- processing and features detection

The preview method for detecting sunspots for W.L. 393 nm by Meftah 2016 has limitations, because a manual threshold has to be used. Moreover, these steps are the most time consuming and the method only applies to 393nm W.L. images, so does not apply to large data.

The method for this study overcomes problems associated with time consumption because it is automated and can be applied on large data for 607nm W.L. It also provides better results than those produced on a 535nm W.L. The method is developed to automatically detect sunspots in 607nm SODISM L1 images and is programmed using MATLAB; it adopts the following steps, shown as Algorithms 1 and 2[6].

**Algorithm 1.** Extraction of the solar limb

[i]   Obtain a clean solar disk without noise and sunspots; this can be achieved by applying a dilation and then an erosion operation, i.e. a closing operation with a structuring element (SE) on an original SODISM image

[ii]  Choose a circular SE of 30 pixels radius (this value was chosen by cross validation, the biggest sunspot in many of images is chosen, and its radius calculated to be 30 pixels, so the SE radius is set to be 30). The sample result is shown in Figure 3.b

[iii] To secure the solar limb, determine the border edges; thus, shrink the solar disk by one pixel (filtered image in Figure 3.b) to produce a smaller image,

[iv]  Then subtract the new image from the filtered image; the result is illustrated in Figure 3.c

[v]   Eliminate CCD noises by utilizing a Kuwahara Filter (refer to Figures 3.d and 3.e).

Apply a binary overlay plugin between the original and solar limb images; it is labelled with a red colour and overlapped on the original image, as shown in Figure 3.f.

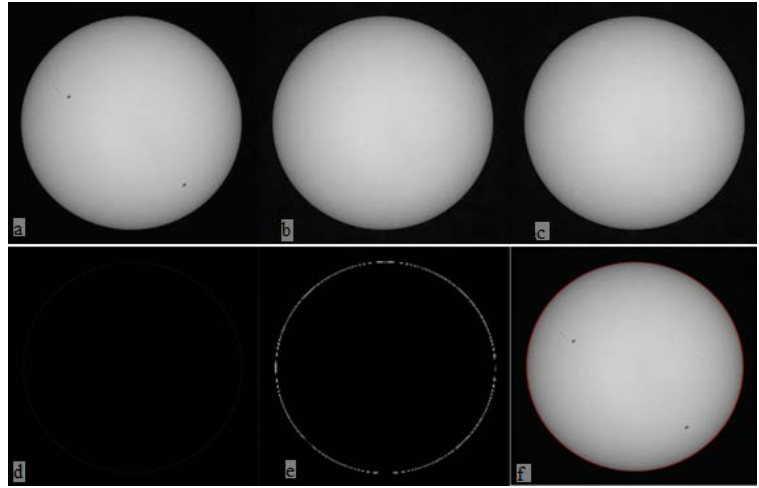

**Figure 3.** (As shown from left to right, top to bottom): (a) the original image; (b) the filtered image; (c) the shrunken image of the solar disk, where the radius is 1 pixel smaller than that in (b); (d) the solar limb shown in a grey image; (e) solar limbs; (f) red colour around disk shows the solar limb label

**4. Detection of Sunspots and verification**

　　The detection of sunspots involves the recognition of sunspots on the solar disk after the solar limb has been extracted. Due to the limited resolution of the data, the sunspot umbra and penumbra are not separated in the SODISM images. The steps outlined in Algorithm 2 enable the identification of sunspots (refer to Figure 4 for the associated images).

**Algorithm 2.** The Detection of Sunspots.

[i]　Process the original image from the PICARD website using the proposed quality enhancement method; the sample output is shown in Figure 4.a

[ii]　Compute the gradient of the sunspot boundaries (refer to Figure 3.a)

[iii]　Fill the holes with a closing operation; this leads to the removal of dark regions surrounded by bright crests in grayscale images.

[iv]　Compute the image obtained in 4.a and the gradient image obtained in [ii] (i.e. the difference between Figures 4.a and 4.b will yield 4.c).

[v]　Separate the sunspot gradient from the noises, as shown in Figure 4©. Many experiments were applied to ascertain a suitable value and an intensity of 15% (in Figure 4.c) was identified; however, due to the solar limb darkening, it was noted that the sunspot's gradient was lower at the solar limb, so the threshold was 10% in the region of the solar disk. Remove the unwanted noises using the Kuwahara Filter. Employ the Extended Min and Max operation as a marker detection to enable segmentation, imposing minimal or maximal on a grayscale image is provided: Figure 4.d shows the sunspot candidate.

[vi]　Acquire sunspots from the candidates, as shown in Figure 4.d. This study considered the candidates as verified sunspots in which the difference between the maximum and minimum grey values of a pixel are greater than 5, and the other regions are ignored.

　　Apply a binary overlay in a red colour and superimpose the original image, (Figure 5 shows the sample result).

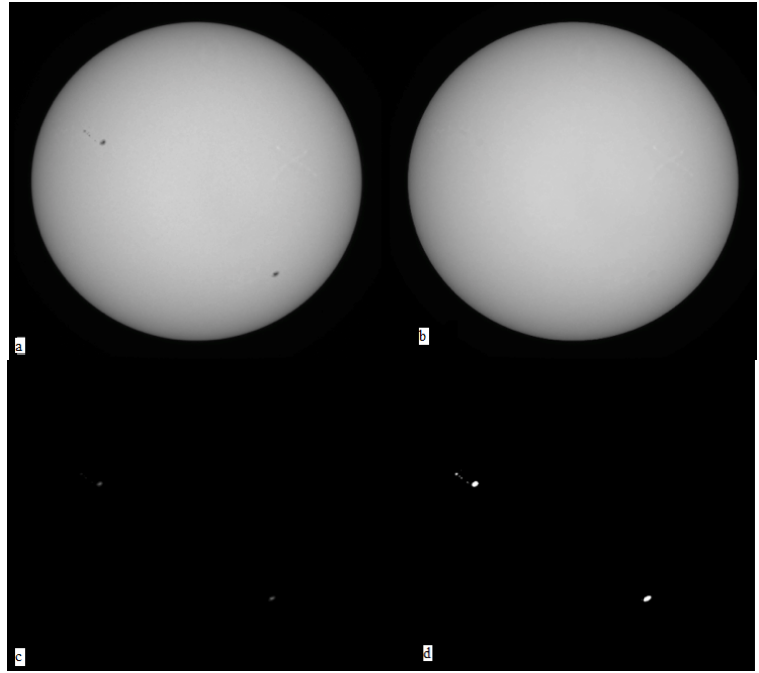

**Figure 4.** (a) The original image disturbed by instrument noises; (b) the filtered image without sunspots (c) The gradient on the image; (d) the binary image showing sunspot candidates

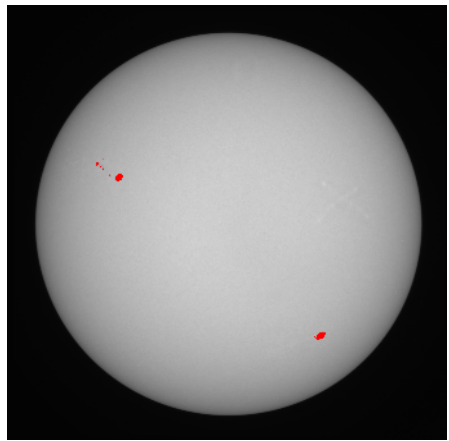

**Figure 4**. The recognised and superimposed sunspots on the original image

**5. Accuracy of the automatic procedure compared with the NOAA catalogue**

The data from the 607nm W.L. from August and September 2010, for both the SODISM automated detection process and NOAA are illustrated in Table 1. The table has five columns: the first shows the date on which the image was captured; the second represents the number of sunspots in the image (manually counted), and the third column shows the number of sunspots detected by the study's automatic pipeline. The fourth column shows the false rejection rate (FRR), i.e. the number of sunspots detected automatically, but not detected by the NOAA catalogue, and finally, the last column is the false acceptance rate (FAR) i.e. the number of sunspots detected by the NOAA catalogue but not by the automatic method. The rationale for computing the FRR and FAR is to evaluate the proposed technique in accordance with the methods of previous research [10]. SOHO images have been used as a reference in order to overcome small sunspots that were missed through manual processing due to the limitation of visibility conditions. The total number of sunspots detected is listed in the last row of the table for both the automated method and the NOAA method respectively. The recognition rate is calculated as follows:

$$= \frac{\text{sum of the automatic method} - \text{sum of the FAR}}{\text{sum of NOAA}} \quad (1)$$

$$= \frac{89 - 7}{84} \times 100\%$$

$$= 97.6\%$$

**Table 1.** The Comparison of automatically Detected Sunspots with NOAA

| Date | SSs No. (automatic method) | Time of SSs in SODISM image | SSs No. (NOAA catalogue) | false rejection rate FRR | false acceptance rate FAR |
|---|---|---|---|---|---|
| 05/08/2010 | 4 | 04:49 | 4 | 0 | 0 |
| 06/08/2010 | 4 | 01:07 | 4 | 0 | 0 |
| 07/08/2010 | 4 | 01:07 | 4 | 0 | 0 |
| 09/08/2010 | 4 | 05:27 | 4 | 0 | 0 |
| 10/08/2010 | 4 | 01:17 | 3 | 1 | 0 |
| 11/08/2010 | 5 | 05:43 | 5 | 0 | 0 |
| 12/08/2010 | 4 | 00:47 | 0 | 4 | 0 |
| 13/08/2010 | 4 | 04:01 | 2 | 2 | 0 |
| 14/08/2010 | 2 | 05:27 | 2 | 0 | 0 |
| 15/08/2010 | 2 | 00:31 | 2 | 0 | 0 |
| 16/08/2010 | 3 | 03:21 | 3 | 0 | 0 |
| 17/08/2010 | 2 | 05:39 | 2 | 0 | 0 |
| 18/08/2010 | 1 | 01:25 | 2 | 0 | 1 |
| 19/08/2010 | 1 | 01:33 | 2 | 0 | 1 |
| 20/08/2010 | 0 | 03:15 | 1 | 0 | 1 |
| 21/08/2010 | 0 | 03:59 | 0 | 0 | 0 |
| 22/08/2010 | 0 | 00:07 | 0 | 0 | 0 |
| 23/08/2010 | 0 | 02:21 | 0 | 0 | 0 |
| 24/08/2010 | 0 | 05:41 | 1 | 0 | 1 |
| 25/08/2010 | 1 | 00:51 | 0 | 1 | 0 |
| 26/08/2010 | 2 | 06:39 | 2 | 0 | 0 |
| 27/08/2010 | 1 | 01:55 | 1 | 0 | 0 |
| 28/08/2010 | 1 | 05:47 | 1 | 0 | 0 |
| 29/08/2010 | 2 | 01:11 | 2 | 0 | 0 |
| 30/08/2010 | 2 | 02:51 | 2 | 0 | 0 |
| 04/09/2010 | 3 | 12:13 | 3 | 0 | 0 |
| 05/09/2010 | 4 | 06:19 | 4 | 0 | 0 |
| 06/09/2010 | 2 | 02:07 | 1 | 1 | 0 |
| 07/09/2010 | 0 | 21:20 | 1 | 0 | 1 |
| 11/09/2010 | 0 | 00:12 | 0 | 0 | 0 |
| 12/09/2010 | 1 | 08:05 | 2 | 0 | 1 |
| 13/09/2010 | 3 | 00:13 | 1 | 2 | 0 |
| 20/09/2010 | 3 | 07:59 | 2 | 1 | 0 |
| 22/09/2010 | 2 | 04:49 | 2 | 0 | 0 |
| 23/09/2010 | 2 | 14:11 | 2 | 0 | 0 |
| 24/09/2010 | 2 | 03:15 | 2 | 0 | 0 |
| 25/09/2010 | 2 | 03:25 | 2 | 0 | 0 |
| 26/09/2010 | 2 | 03:25 | 3 | 0 | 1 |
| 27/09/2010 | 3 | 03:27 | 3 | 0 | 0 |
| 29/09/2010 | 3 | 04:21 | 3 | 0 | 0 |
| 30/09/2010 | 4 | 00:47 | 4 | 0 | 0 |
| **Total** | **89** |  | **84** | **12** | **7** |

## 6. Filling factors

The filling factor is calculated as a function of the radial position on the sun disk. Thus, the calculated filling factors for a particular feature reflect the fraction of the solar disk covered by the feature, to which synthetic spectra references are assigned [14]. Eleven concentric rings divide the solar disk; these start with an inner radius (RI), and conclude with an outer radius (RO). Figure 6 shows the filling factor represented on solar disks and Figure 7 shows the filling factor coverage for sunspots.

**Table 1.** Relative radius values

| Index | Inner radius (relative radius) | Outer radius (relative radius) |
|---|---|---|
| 1 | 0.00 | 0.07 |
| 2 | 0.07 | 0.16 |
| 3 | 0.16 | 0.25 |
| 4 | 0.25 | 0.35 |
| 5 | 0.35 | 0.45 |
| 6 | 0.45 | 0.55 |
| 7 | 0.55 | 0.65 |
| 8 | 0.65 | 0.75 |
| 9 | 0.75 | 0.85 |
| 10 | 0.85 | 0.95 |
| 11 | 0.95 | 1.05 |

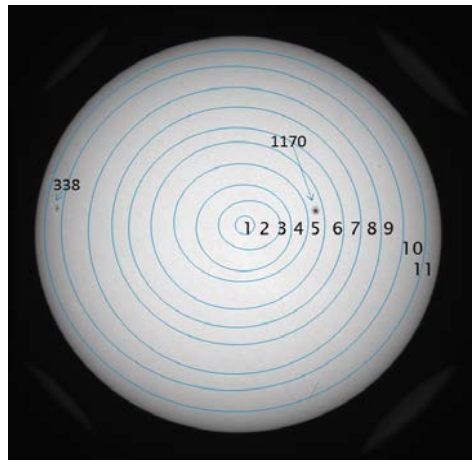

**Figure 5.** An example for filling factor rings

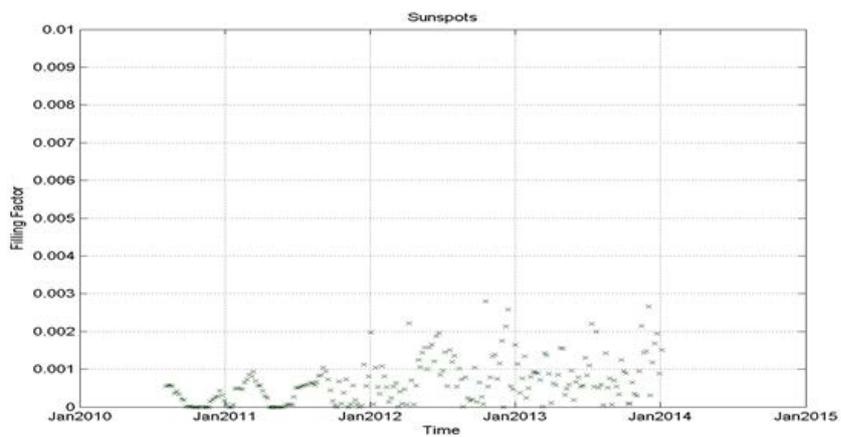

**Figure 6.** The filling factors (area coverage) for sunspots

The data obtained for the 607nm W.L. were collected from 22th September 2010 to 1st January 2014. The filling factor for SOHO's images have been calculated and compared with SODISM over the same period (i.e. September, October, November and December 2010). The correlation coefficient was 99%, which reflects that the method gives excellent results for the 607nm W.L. Figure 8 shows the comparison with the SOHO satellite.

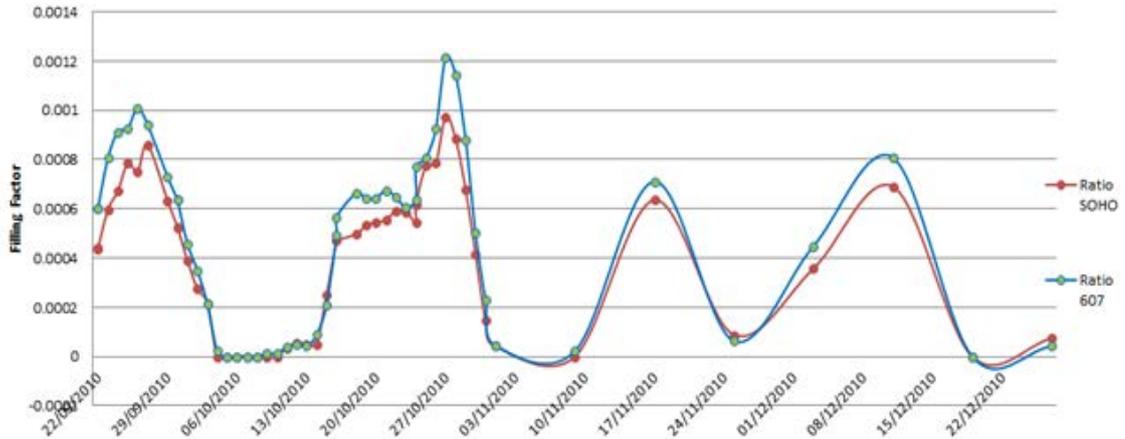

**Figure 7.** Filling factor calculation of sunspot from SOHO and SODISM images W.L. 607 nm from 22th September 2010 to 25 December 2010[6]

**7. Filling factor calculations**

To catalogue these procedures, the results have been summarised in tabular form to provide a single snapshot of the completed work. Parametric and non-parametric statistics are used to describe the data. While parametric statistics make the assumption that the data is drawn from a population with a normal distribution, non-parametric tests make no such assumption. Hence, working with both provides robustness which counters the shortcomings of each statistical test; thus, the results should complement each other. The parametric measures employed are the three statistical measures of mean, variance and skewedness. In comparison, medians and interquartile ranges (IQRs) are used as non-parametric measures. Since skewedness gives information about the normality of the distribution, it could be expected that the results of the first and second statistical measures (mean and variance) would be reliable if the third (higher) measure indicates normality. On the other hand, an individual may rely on information provided by the non-parametric statistics if the third order measure indicates non-normality. As a rule of thumb, researchers assume that a value of skewedness above ±2 implies a distribution that is highly skewed and hence not normally distributed [15]. Furthermore, the IQR can be used to compute the number of outliers in the data[16] using the following equation:

$$\text{for } x \in \text{observations} \qquad (2)$$

Where x is an outlier if it satisfies any of the two conditions.

$$x \begin{cases} < q_1 - 1.50 \text{ IQR} \\ > q_3 + 1.5 \text{ IQR} \end{cases} \qquad (3)$$

Where $q_1$ and $q_3$ are the first and third quartiles respectively. Thus, the proposed cataloguing table will contain the information listed in table 3.

Table 2: Proposed cataloguing table

| Total number of samples n=206 | |
|---|---|
| Test | Result |
| μ(Total Sunspot) | 1876.6 |
| σ(Total Sunspot) | 1658.8 |
| skew(Total Sunspot) | 1.1143 |
| Med(Total Sunspot) | 1573.0 |
| IQR(Total Sunspot) | 2106.8 |
| Total Sunspot Outliers | 7 |
| μ(Sunspot Ratio) | 0.000662 |
| σ(Sunspot Ratio) | 0.000585 |
| skew(Sunspot Ratio) | 1.1284 |
| Med(Sunspot Ratio) | 0.000556 |
| IQR(Sunspot Ratio) | 0.000755 |
| Sunspot Ratio Outliers | 7 |

From the results presented in Table 3, the skewedness measure indicates that the data are normally distributed and thus, the parametric scores are reliable. It is therefore possible to report that the extensive evaluation conducted on 206 SODISM data images achieved a mean sunspot region of 1876.6 and a corresponding average sunspot ratio of 0.000662. It can further be deduced that any image picked from the dataset is expected to have a FF signature of ±3 standard deviation; moreover, the value of σ can be conveniently retrieved from Table 3. Furthermore, the IQR-based outlier detection has revealed that there are seven images within the data that lie outside the normal distribution. Hence, as with any other real-world data, it is expected that these few outliers (≈ 0.032% of the dataset) will exhibit some irregularities.

## 8. Concluding discussions

The proposed segmentation method has been applied to the entire downloaded 607nm image data in order to detect sunspots and calculate their filling factors. Moreover, a comparison with the NOAA catalogue has been conducted. Figure 7 shows filling factor coverage from October 2010 until the end of life for the Picard satellite, which was on the 1st January 2014. Moreover, Figure 8 shows a comparison between the filling factors calculated for the SODISM 607nm images and the MDI intensitygram images from the SOHO satellite over a similar period (i.e. from 22th September, 2010 to 24th December, 2010). The comparison between the two values, or measure of dependence between the two quantities, is calculated as Pearson's correlation coefficient, which is 99% between SOHO and SODISM. Moreover, from Table 1, the recognition rate for the proposed method is approximately 98%.

It is possible to use suitable automated methods for detecting sunspots on SODISM images, despite the image degradation throughout the lifetime of PICARD. The biggest advantage is the reduction in time consumption. There are only a few methods applied to segmented SODISM images; the first was applied by Meftah *et al.* [11] to 393nm W.L. but needed manual interaction to optimize the threshold, which was calculated using the Otsu method [7]. The second method developed by Ahmed *et al.* shows a strong correlation coefficient of 98% [11] between the SODISM and SOHO images. This was only slightly less than that achieved in the same period in the third method by Alasta *et al.*, which successfully detected sunspots on 535nm W.L. images over the lifetime of PICARD, and then calculated the filling factors. Furthermore, a comparison of sunspot filling factors between SOHO and SODISM images indicate a very close match over the early period when both are available, achieving a correlation coefficient of 98.5% [7].

The completely automated method applied to the W.L. 607nm; makes it easily applicable to large data images. The 0.99 correlation coefficient reflects excellent results. The results in Figure 8 show that the filling factors for SODISM and SOHO are slightly different in amplitude despite mostly changing instep. This is most evident between 22 September 2010 and 29 September 2010 when there is a somewhat lower correlation coefficient ($\approx$0.95). Nevertheless, this is an improvement on the results from the previous method at a 535nm W.L. over the same period, which showed a correlation 0.92. This lower correlation could be related to the fact that the SOHO data corresponds to a different wavelength (676.8nm) than the SODISM images. However, this is the first automated method to achieve 0.99 correlation between SOHO and SODISM. Table 3 illustrates a key technique for identifying regions of interest; image segmentation was also explored, investigated and deployed. An evaluation and comparison of the results from similar works was conducted. In general, the system developed and described in this paper has proven to be promising, in that, out of 89 sunspots, it automatically detected around 98%, which is comparable with the NOAA catalogue of sunspots. Moreover, a statistical means has been presented to represent SODISM data, which, as stated earlier, is a useful innovation. The afore-defined metrics present researchers with a new method of summarising hundreds of experiments in a concise and easily understood manner. The catalogue can then be used as a template to conduct comparisons. Should researchers embark on work similar to the one presented in this paper, it is hoped that they will use this novel procedure to present their findings as a means for comparison.

**References**

[1] M. Meftah, T. Corbard, A. Hauchecorne, A. Irbah, P. Boumier, A. Chevalier, W. Schmutz, R. Ikhlef, F. Morand, C. Renaud, J.-F. Hochedez, G. Cessateur, S. Turck-Chièze, D. Salabert, M. Rouzé, M. van Ruymbeke, P. Zhu, S. Kholikov, S. Koller, C. Conscience, S. Dewitte, L. Damé, and D. Djafer, "Main results of the PICARD mission," no. July, p. 99040Z, 2016. Available: http://dx.doi.org/10.1117/12.2232027.

[2] A. F. Alasta, A. Algamudi, R. Qahwaji, S. Ipson, and T. A. Nagern, "Automatic sunspots detection on SODISM solar images," in 7th International Conference on Innovative Computing Technology, INTECH 2017, 2017, pp. 115–119. Available: http://ieeexplore.ieee.org/document/8102429/.

[3] M. Meftah, a Hauchecorne, T. Corbard, E. Bertran, M. Chaigneau, and M. Meissonnier, "PICARD SODISM, a space telescope to study the Sun from the middle ultraviolet to the near infrared," no. January, pp. 1–38, 2014. Available: https://link.springer.com/article/10.1007%2Fs11207-013-0373-x.

[4] M. Meftah, A. Irbah, A. Hauchecorne, and J.-F. Hochedez, "PICARD payload thermal control system and general impact of the space environment on astronomical observations," no. May 2013, p. 87390B, 2013. Available: http://proceedings.spiedigitallibrary.org/proceeding.aspx?doi=10.1117/12.2010178

[5] R. Qahwaji and T. Colak, "Automatic detection and verification of solar features," Int. J. Imaging Syst. Technol., vol. 15, pp. 199–210, 2005. Available: https://onlinelibrary.wiley.com/doi/pdf/10.1002/ima.20053

[6] A. F. Alasta, A. Algamudi, R. Qahwaji, S. Ipson, and T. A. Nagern, "Automatic sunspots detection on SODISM solar images," in 2017 Seventh International Conference on Innovative Computing Technology (INTECH), 2017, pp. 115–119.

[7] J. J. Curto, M. Blanca, and E. Martínez, "Automatic sunspots detection on full-disk solar images using mathematical morphology", Sol. Phys., vol. 250, no. 2, pp. 411–429, 2008.

[8] S. Zhang, H. Yang, and L. Singh, "Increased information leakage from text", CEUR Workshop Proc., vol. 1225, no. 2003, pp. 41–42, 2014.

[9] V. Zharkova, S. Ipson, A. Benkhalil, and S. Zharkov, "Feature recognition in solar images", Artif. Intell. Rev., vol. 23, no. 3, pp. 209–266, 2005.

[10] O. A. Rami Qahwaji, Stan Ipson, "SOLID D3.5 Filling factors catalogue for PICARD images SOLID", First Eur. Compr. Sol. Irradiance Data Exploit., pp. 1–8, 2015.

[11] A. F. Alasta, A. Algamudi, R. Qahwaji, S. Ipson, A. Hauchecorne, and M. Meftah, "New method of enhancement using wavelet transforms applied to SODISM telescope", Advances in Space Research, Aug. 2018. Available: https://www.sciencedirect.com/science/article/pii/S0273117718306112?via%3Dihub.

[12] A. F. Alasta, "Using Remote Sensing data to identify iron deposits in central western Libya," International Conference on Emerging Trends in Computer and Image Processing (ICETCIP'2011), Bangkok. Available: http://psrcentre.org/images/extraimages/122.%201211924.pdf.

[13] A. F. Alasta, A. Algamudi, and S. Ipson, "Identification of Sunspots on SODISM Full-Disk Solar Images", IEEE Int. Conf. Comput. Electron. & Commun. Eng. 2018 (iCCECE '18) Univ. Essex, Southend, UK, pp. 23–28, 2018.

[14] O. Ashamari, R. Qahwaji, S. Ipson, M. Schöll, O. Nibouche, and M. Haberreiter, "Identification of photospheric activity features from SOHO/MDI data using the ASAP tool", J. Sp. Weather Sp. Clim., vol. 5, p. A15, 2015. Available: http://arxiv.org/abs/1505.02036.

[15] F. J. Gravetter and L. B. Wallnau, Statistics for the behavioral sciences. Cengage Learning, 2016.

[16] P. J. Rousseeuw and M. Hubert, "Robust statistics for outlier detection", Wiley Interdiscip. Rev. Data Min. Knowl. Discov, vol. 1, no. 1, pp. 73–79, 2011.